\documentclass{aa}
\usepackage{epsfig}
\usepackage{psfig}
\usepackage{graphicx}
\usepackage{txfonts}

\usepackage{lscape}

\newcommand{\PSR}{PSR~J1509$-$5850}
\titlerunning{The pulsar wind nebula associated with \PSR}
\begin{document}
\title{Radio and X-ray nebulae associated with \PSR}

\author{C. Y. Hui \and W. Becker}   
\date{Received 11 April 2007 / Accepted 10 May 2007}
\institute{Max-Planck Institut f\"ur Extraterrestrische Physik, 
          Giessenbachstrasse 1, 85741 Garching bei M\"unchen, Germany}

\abstract{We have discovered a long radio trail at 843 MHz which is 
apparently associated with middle age pulsar \PSR. The radio trail has 
a length of $\sim 7$ arcmin. In X-rays, Chandra observation of \PSR\ 
reveals an associated X-ray trail which extends in the same orientation 
as the radio trail. Moreover, two clumpy structures are observed along the 
radio trail. The larger one is proposed to be the supernova remnant (SNR)
candidate MSC 319.9-0.7. Faint X-ray enhancement at the position of 
the SNR candidate is found in the Chandra data. 
 
\keywords{pulsars: individual (\PSR )---stars: neutron---radio, X-rays: stars}}

\maketitle

\section{Introduction}
 It is generally believed that a significant fraction of the rotational energy 
 of a pulsar leaves the magnetosphere in the form of a magnetized pulsar wind 
 consisting of electromagnetic radiation and high energy particles. In 
 view of this, it is energetically important to study the physical properties 
 of this wind. When the relativistic wind particles interact with the shocked 
 interstellar medium, the charged particles will be accelerated in the shock
 and hence synchrotron radiation from radio to X-ray is generated. 
 In order to obtain a better understanding of the interaction 
 nature, multiwavelength studies of the pulsar wind 
 nebulae are deeply needed. X-ray and radio observations have 
 recently revealed a number of pulsar wind nebulae. 
%TeV observations have suggested the existence of PWN around several pulsars.
 However, there is only a handful of shocked emission detected in 
 both the X-ray and radio regimes (c.f.~see Hui \& Becker 2006 and 
 references therein). 

 \PSR\ was discovered by Manchester et al.~(2001) in the Parkes Multibeam 
 Pulsar Survey. The pulsar has a rotation period of $P=88.9$ ms and a period 
 derivative of $\dot{P}=9.17\times 10^{-15}$ s s$^{-1}$. These spin parameters 
 imply a characteristic age of $1.54\times 10^{5}$ yrs, a dipole surface 
 magnetic field of $B_{\perp}=9.14\times 10^{11}$ G and a spin-down luminosity 
 of $5.1\times 10^{35}$ ergs s$^{-1}$ (c.f.~Table 1). The radio dispersion 
 measure yields a distance of about 3.81 kpc based on the galactic free
 electron model of Taylor \& Cordes (1993). Using the model of  Cordes \&
 Lazio (2002) the dispersion measure based distance is estimated to be
 2.56 kpc. The proper motion of this pulsar is not yet known. Recently, 
 a brief X-ray study of the field of \PSR\ was presented by Kargaltsev et 
 al.~(2006). The authors have reported that a trail-like pulsar wind nebula 
 associated with \PSR\ was observed in a Chandra observation. The X-ray nebula 
 is found to be extended in the south-west direction. 

 In this paper we report on the discovery of a possible radio counterpart 
 of the X-ray trail associated with \PSR\ and provide a detailed X-ray 
 analysis of the trail. In \S2 we describe the observations and the data 
 analysis and in \S3 we summarize and discuss our results. 

 \begin{table}
 \centering
 \caption{Pulsar parameters of \PSR\ (from Manchester et al.~2005) }
 \begin{tabular}{lc}
 \hline\hline
                      &  \\
 \hline
 Right Ascension (J2000)  & $15^{\rm h} 09^{\rm m} 27.13^{\rm s}$ \\
 Declination (J2000)          & $-58^\circ\; 50'\; 56.1"$  \\
 Pulsar Period, $P$ (s)        & 0.088921760489  \\
 Period derivative $\dot{P}$ ($10^{-15}$ s s$^{-1}$) & 9.1698 \\
 Age ($10^{5}$ yrs)          & 1.54 \\
 Surface dipole magnetic field ($10^{12}$ G) & 0.914 \\
 Epoch of Period (MJD)      &  51463 \\
 Dispersion Measure (pc cm$^{-3}$)  & 137.7 \\
 Dispersion based distance (kpc)   & $\sim 2.6 - 3.8$  \\
 Spin-down Luminosity ($10^{35}$) ergs s$^{-1}$ & 5.1 \\
 \hline
 \end{tabular}
 \end{table}

\section{Observations and data analysis}
\PSR\ was observed with Chandra in 2003 February 9$-$10 (Obs ID: 3513) 
with the Advanced CCD Imaging Spectrometer (ACIS). The pulsar is located 
on the back-illuminated (BI) ACIS-S3 chip which has a superior quantum 
efficiency among the spectroscopic array. Standard processed level-2 data 
were used. The effective exposure is about 40 ks. 

Chandra observation has revealed an X-ray trail associated with \PSR. 
The signal-to-noise ratios for the pulsar and the trail are found to 
be $\sim 19$ and $\sim 3$ in $0.5-8$ keV respectively. 
The X-ray image of the $4\times4$ arcmin field near to \PSR\ is shown 
in Figure 1. The binning factor of the image is 0.5 arcsec. Adaptive 
smoothing with a Gaussian kernel of $\sigma < 3$ arcsec has been 
applied to the image. The trail appears to have a length of $\sim 2$ 
arcmin. From a Digitized Sky Survey (DSS) image, 25 bright field stars 
are found in the field of view of Figure 1. We subsequently identified 
the magnitudes of these stars from the USNO-A2.0 catalog (Monet et al.~1998), 
which are within the range of $B\sim 10-18$. Their positions are plotted 
as white circles in Figure 1. 

\begin{figure}
\psfig{figure=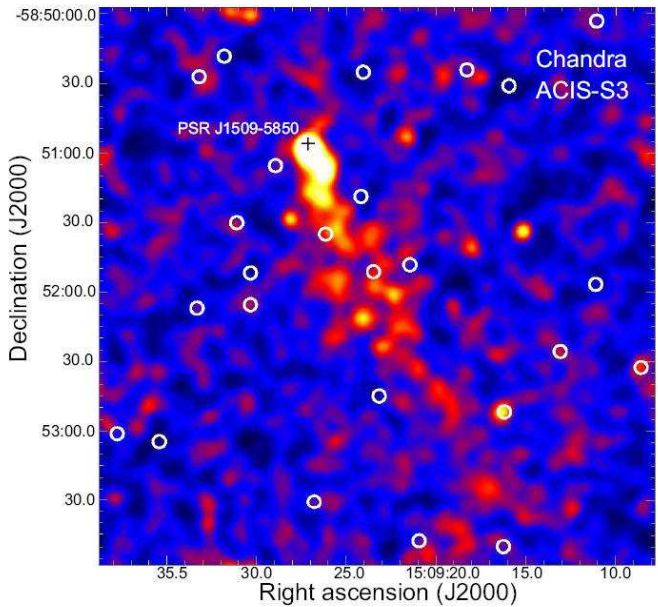,width=9cm,clip=}
\caption{Chandra's $4\times4$ arcmin view of \PSR\ and its X-ray trail 
in the energy band 0.3$-$8 keV. The pulsar position is indicated
by the black cross. The white circles indicate the positions of 
field stars identified in the DSS image.} 
\end{figure}

For the spectral analysis, we extracted the spectrum of \PSR\ from a circle 
of 4 arcsec radius (encircled energy$\sim$99\%) centered on the pulsar position. 
To minimize the possible contamination from the field stars,  the spectrum from 
the trail was extracted within a box of $25\times 95$ arcsec, oriented along 
the direction of the trail emission. Even with this consideration, there are 
still two stars with magnitude $B=17$ and $B=16.4$ located on the trail
(cf.~Fig.~1) and unavoidably lie in the extraction region. Without the knowledge 
of the extinctions, we are not able to estimate the possible contribution in 
X-ray from these two stars. The background spectra were extracted from the 
low count regions nearby. After background subtraction, there are $\sim 100$ 
net counts and $\sim 270$ net counts extracted from the pulsar and the trail 
in $0.5-8$ keV respectively. Response files were computed by using the CIAO 
tools MKRMF and MKARF. The spectra were dynamically binned so as to have at 
least 10 counts per bin for the pulsar and 30 counts per bin for the trail. 
All the spectral fitting were performed in the energy range of $0.5-8$ keV 
by using XSPEC 11.3.1. The degradation of the ACIS quantum efficiency was 
corrected by XSPEC model ACISABS. All the quoted errors are $1-\sigma$ and 
were computed for 1 parameter in interest.

For the X-ray emission from \PSR, we found that it can be modeled with an 
absorbed power-law fairly well ($\chi_{\nu}$=0.68 for 8 D.O.F.). This model 
yields a column density of $N_{H}=8.0^{+2.3}_{-2.1}\times 10^{21}$ cm$^{-2}$, 
a photon index of $\Gamma=1.0^{+0.2}_{-0.3}$ and a normalization at 1 keV of 
$5.1^{+1.3}_{-1.6}\times 10^{-6}$ photons keV$^{-1}$ cm$^{-2}$ s$^{-1}$. The 
best-fitted model results in an unabsorbed flux of $f_{X}=5.9\times 10^{-14}$ 
ergs cm$^{-2}$ s$^{-1}$ in the energy range of $0.5-8$ keV. The dispersion 
based distance implies a luminosity of $L_{X}=4.8\times 10^{31}$ and 
$1.0\times 10^{32}$ erg s$^{-1}$ for $d$=2.6 and 3.8 kpc respectively.  
Although a blackbody model can give a compatible goodness-of-fit ($\chi_{\nu}$
=0.82 for 8 D.O.F.), it infers a rather high temperature ($T\sim 1.7\times 10^{7}$ K) 
and a small projected blackbody radius ($R\sim 10$ m). We hence regard this model 
as not physically reasonable to describe the X-ray spectrum of \PSR. We note 
that the characteristic age indicates that \PSR\ belongs to the class of middle-aged 
pulsars. Their spectra typically consist of a soft thermal component, a harder thermal 
component from the heated polar caps as well as contribution from the non-thermal 
emission (cf.~Becker \& Aschenbach 2002). However, the small number of collected photons 
and the high column density does not support any fitting with multicomponent models. 

We have tested the hypothesis that the trail emission is originated from the 
interaction of pulsar wind and ISM by fitting an absorbed power-law model to 
the trail spectrum. The model yields an acceptable goodness-of-fit 
($\chi_{\nu}$=0.73 for 9 D.O.F.). The best fitting spectral model is displayed 
in Figure 2. This model yield a column density of $N_{H}=8.2^{+9.3}_{-3.7}
\times 10^{21}$ cm$^{-2}$, a photon index of $\Gamma=1.3^{+0.8}_{-0.4}$ and a 
normalization at 1 keV of $1.9^{+4.3}_{-1.9}\times 10^{-5}$ photons 
keV$^{-1}$ cm$^{-2}$ s$^{-1}$. We note that the column density agrees 
with that inferred from the pulsar spectrum. The unabsorbed flux deduced 
for the best-fitted model parameters are $f_{X}=1.6\times 10^{-13}$ erg s$^{-1}$ cm$^{-2}$ 
in the energy range of $0.5-8$ keV. The dispersion based distance 
implies a luminosity of $L_{X}=1.3\times 10^{32}$ and $2.7\times 10^{32}$ erg s$^{-1}$   
for $d$=2.6 and 3.8 kpc, respectively. 

\begin{figure}
 \centerline{\psfig{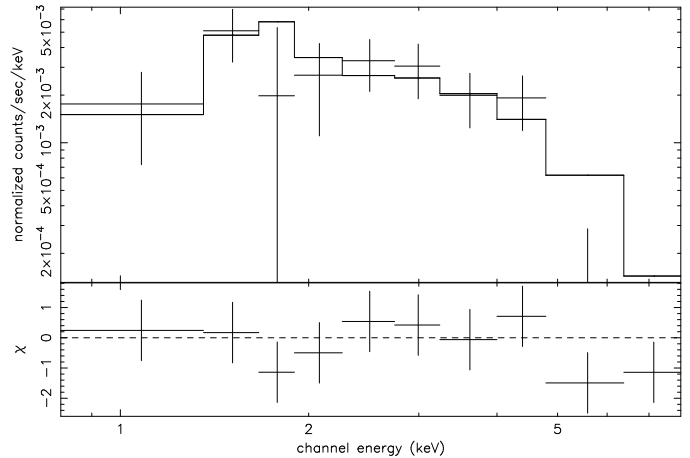}}
 \caption{Energy spectrum of the X-ray trail of \PSR\, as observed with
 the Chandra ACIS-S3 detector and fitted to an absorbed power-law model
 ({\it upper panel}) and contribution to the $\chi^{2}$ fit statistic
 ({\it lower panel}).}
 \end{figure} 

We have searched for a possible radio counterpart for the X-ray nebula with the 
Sydney University Molonglo Sky Survey data (SUMSS) (Bock et al.~1999). We have 
discovered a long radio trail apparently associated with \PSR. The radio image 
of the $11\times11$ arcmin field near to \PSR\ is displayed in Figure 3. The 
radio feature has a length of $\sim7$ arcmin. Radio contours were calculated 
at the levels of 7, 23, 39, 54 and 70 mJy/beam. These contours were 
overlaid on the image of the Chandra ACIS-S3 chip in Figure 4. 
It is interesting to note that the radio trail starts exactly from the position 
of \PSR\ and has the same orientation as that of the X-ray trail. These facts 
support the interpetation that this extended radio feature is the radio counterpart 
of the X-ray trail and is indeed physically associated with \PSR.

\begin{figure}
\psfig{figure=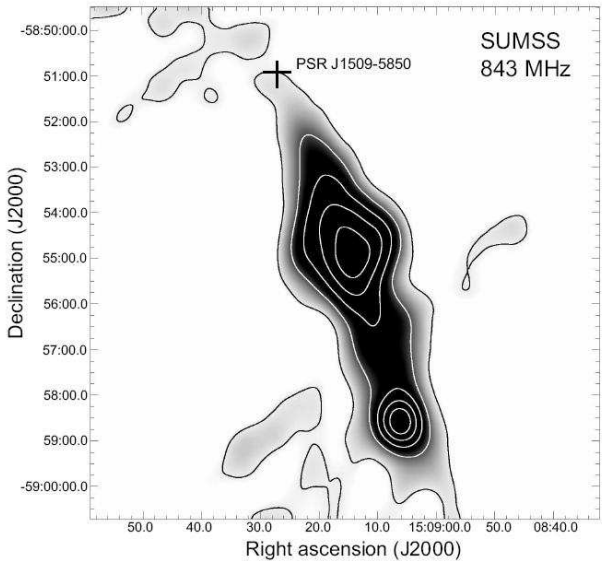,width=8.4cm,clip=}
\caption{The 843 MHz SUMSS image of a field of $11\times11$ arcmin around 
 PSR~J1509-5850. The pulsar position is  indicated by the black cross. 
 The $\sim7$ arcmin long radio feature is found to have the same orientation 
 as the X-ray trail. The contour levels are 7, 23, 39, 54 and 70 mJy/beam. 
 Two clumps are observed along the trail. The larger clump, near to the center 
 of this image, on the trail is identified as  the SNR candidate MSC 319.9-0.7.}
\end{figure}

There are two clumpy structures observed along the radio trail (see Figure 3). 
The northern clump has its emission center at RA=$15^{\rm h} 09^{\rm m} 14.35^{\rm s}$, 
Dec=$-58^\circ\; 54'\; 50.7"$ (J2000) with a radius of $\sim 1.5$ arcmin. 
The southern clump has its emission center at 
RA=$15^{\rm h} 09^{\rm m} 06.33^{\rm s}$, 
Dec=$-58^\circ\; 58'\; 34.7"$ (J2000) with a radius of $\sim 1$ arcmin. 
While the southern clump is unidentified in SIMBAD and NED, the northern 
clump, which locates $\sim 4$ arcmin away from \PSR, has been proposed to 
be a supernova remnant candidate MSC 319.9-0.7 (Whiteoak \& Green 1996). 
Comparing the X-ray and the radio data in Figure 4, we found that there 
is some faint X-ray emission near to the location of MSC 319.9-0.7. The 
emission does not appear to be the continuation of the trail associated 
with \PSR. It cannot be excluded that this faint emission is related to 
MSC 319.9-0.7. However, the limited photon statistics does not allow any 
final conclusion. 

\begin{figure}
\psfig{figure=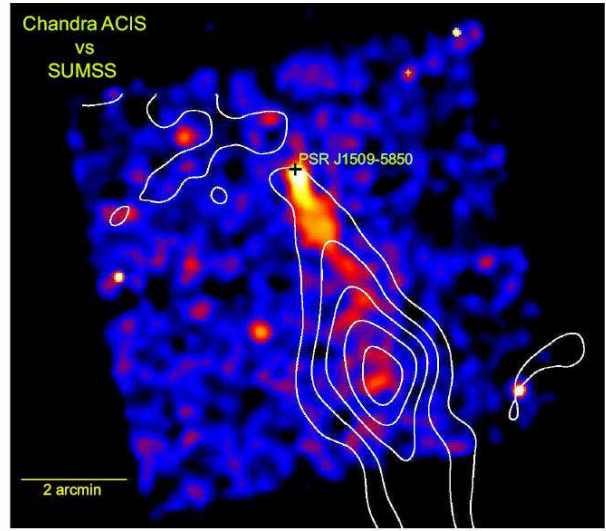,width=8.4cm,clip=}
\caption{The X-ray image of the Chandra ACIS-S3 chip with the radio 
contour lines from SUMSS data (cf.~Figure 3) overlaid. The X-ray 
image is binned with a pixel size of 2.5 arcsec and adaptively 
smoothed with a Gaussian kernel of $\sigma<7.5$ arcsec. We 
note that there is a faint X-ray feature near to the location of the 
SNR candidate MSC 319.9-0.7. Top is north and left is east.}
\end{figure}

\section{Discussion \& Conclusion}
In this paper, we report the detection of a possible radio counterpart 
of the X-ray trail associated with \PSR\ and present a first detailed 
X-ray study of the X-ray trail. Apart from the radio trail, we have found 
that there are two clumpy structures located on the trail. While the 
smaller one is still unidentified, the larger one, which is located $\sim 4$ 
arcmin away from \PSR, is identified as a SNR candidate MSC 319.9-0.7.

Despite the proximity of MSC 319.9-0.7, it seems unlikely that it is the 
birth place of \PSR. Assuming this shell-like SNR candidate is in a Sedov 
phase, the radius of the shocked shell emission can be estimated by (Culhane 1977):

\begin{equation}
R_{s}=2.15\times 10^{-11}\left(\frac{E}{n}\right)^{\frac{1}{5}}
t^{\frac{2}{5}} {\rm pc}
\end{equation}

\noindent where $t$, $E$ and $n$ are the time after the explosion in units of years, 
the released kinetic energy in units of ergs and the ISM number density in units of 
cm$^{-3}$ respectively. Taking the typical values of $E=10^{51}$ ergs and $n=1$ cm$^{-3}$ 
and $t$ to be the characteristic age of \PSR, we estimate that a SNR associated with 
\PSR\ should have a radius of $R_s\sim 40$ pc. However, 
MSC 319.9-0.7 only has a radius of $\sim 1.1-1.7$ pc for $d=2.6-3.8$ kpc. 
Thus, the discrepancy between the expected $R_s$ and the observed value which
with a factor of $\sim 30$ is not likely to be reconciled by the uncertainty 
of the dispersion based distance.  On the other hand, the characteristic age of the 
pulsar can be older than its actual age if its inital spin period was close
to its current period. However, to reconcile such discrepancy would require 
$t$ to be smaller by a factor of $\sim 4000$ which is not likely. Moreover, 
associating MSC 319.9-0.7 with \PSR\ would leave the origin of the southern part of 
the radio trail unexplained. Thus, with the current knowledge of parameters  
it seems most reasonable for us to interpret MSC 319.9-0.7 as a background
source. 

Following the discussion in Hui \& Becker (2006), we apply a simple one zone 
model (Chevalier 2000; Cheng, Tamm, \& Wang 2004) to model the X-ray emission 
properties of the pulsar wind nebula. Since the proper motion of \PSR\ is not 
yet known, we assume the pulsar is in supersonic motion on the basis that the 
nebula resembles a bow-shock morphology. For the supersonic motion, the 
termination shock radius $R_{ts}$ is determined by the balance of the ram 
pressure between the relativistic pulsar wind particles and the ISM at the 
head of the shock (cf.~Cheng et al. 2004):

\begin{equation}
R_{ts}\simeq\left(\frac{\dot{E}}{2\pi\rho_{ISM}v_{p}^{2}c}\right)^{1/2}
\sim 3\times 10^{16}\dot{E}_{34}^{1/2}n^{-1/2}v_{p,100}^{-1} {\rm cm}
\end{equation}

\noindent where $v_{p,100}$ is the velocity of the pulsar in units of 100 km s$^{-1}$, 
$\dot{E}_{34}$ is the spin-down luminosity of the pulsar in units of $10^{34}$ 
erg s$^{-1}$, and $n$ is the number density of the ISM in units of cm$^{-3}$. 
In all the following estimation, we assume \PSR\ has a transverse velocity 
comparable to the average velocity, $\sim 250$ km s$^{-1}$, of ordinary 
radio pulsars (Hobbs et al. 2005). For a ISM density of 1 cm$^{-3}$, equation (2)  
implies a termination radius of $R_{ts}\sim 8.6\times 10^{16}$ cm. 

The X-ray trail is found to be $\sim 2$ arcmin long. For the dispersion 
based distance in the range of $\sim 2.6-3.8$ kpc, the trail has a length 
of $l\sim (4.7-6.8)\times 10^{18}$ cm. 
For the assumed pulsar velocity of $\sim 250$ km s$^{-1}$, the timescale for the 
passage of the pulsar over the length of its X-ray trail, $t_{\rm flow}$, is 
estimated to be $\sim 6000-8600$ years. The magnetic field in the shocked region can 
be estimated by assuming $t_{\rm flow}$ to be comparable to the synchrotron cooling 
timescale of electrons:

\begin{equation}
\tau_{\rm syn}=\frac{6\pi m_{e}c}{\gamma\sigma_{T}B^{2}}
%\tau_{\rm syn}=\left(\frac{6\pi m_{e}c}{\sigma_{\rm T}}\right)
%\left(\frac{3eh}{4\pi m_{e}c}\right)^{\frac{1}{2}}
%\left(h\nu\right)^{-\frac{1}{2}}B^{-\frac{3}{2}}
\simeq 10^{5}\left(\frac{h\nu}{keV}\right)^{-\frac{1}{2}}
B_{\mu G}^{-\frac{3}{2}} {\rm yrs}
\end{equation}

\noindent where $\gamma$ is the Lorentz factor of the wind, taken to be 
$10^{6}$ (cf. Cheng et al. 2004), $\sigma_{T}$ is the Thompson cross 
section, and $B_{\mu G}$ is the 
magnetic field in the shocked region in unit of micro gauss. The inferred 
magnetic field in the shocked region is $\sim 5-7$ $\mu G$. For comparison, 
the magnetic field strength in the ISM is estimated to be $\sim 2-6$ $\mu G$ 
(cf. Beck et al. 2003, and references therein). 

The X-ray luminosity and spectral index depend on the inequality between the 
characteristic observed frequency $\nu_{X}^{\rm obs}$ and the electron synchrotron 
cooling frequency $\nu_{\rm c}$ (see Chevalier 2000 and references therein):

\begin{equation}
\nu_{\rm c}=\frac{18\pi em_{e}c}{\sigma_{T}^{2}\tau^{2}_{\rm syn}B^{3}}
\end{equation}

\noindent which is estimated to be $\nu_{\rm c}=(1.3-1.8)\times 10^{17}$ Hz. Since in 
general $\nu_{X}^{\rm obs}>\nu_{c}$, this suggests the X-ray emission 
is in a fast cooling regime. Electrons with the energy distribution, 
$N\left(\gamma\right)\propto\gamma^{-p}$, are able to radiate their energy 
in the trail with photon index $\alpha=(p+2)/2$. The index $p$ due to shock 
acceleration typically lies between 2 and 3 (cf. Cheng et al. 2004 and reference 
therein). This would result in a photon index $\alpha\simeq 2.0-2.5$. 
In view of the large error of the observed photon index $\Gamma=1.3^{+0.8}_{-0.4}$, 
we cannot firmly conclude the emission scenario simply based on the photon index.
We note that the photon index can still be possibly in the fast cooling regime 
within the $1\sigma$ uncertainty. 
With this consideration and $\nu_{X}^{\rm obs}>\nu_{c}$, we adopted the fast 
cooling scenario in the following discussion. With the assumed value $p=2.2$, 
the calculated photon index $\alpha=2.1$ which is marginally within the 
$1\sigma$ uncertainty of the observed value. 

In a fast cooling regime, the luminosity per unit frequency is given by 
(cf. Cheng et al. 2004):

\begin{equation}
L_{\nu}=\frac{1}{2}\left(\frac{p-2}{p-1}\right)^{p-1}
\left(\frac{6e^{2}}{4\pi^{2}m_{e}c^{3}}\right)^{\frac{p-2}{4}}
\epsilon_{e}^{p-1}\epsilon_{B}^{\frac{p-2}{4}}\gamma^{p-2}
R_{ts}^{-\frac{p-2}{2}}\dot{E}^{\frac{p+2}{4}}\nu^{-\frac{p}{2}}
\end{equation}

Assuming the energy equipartion between the electron and proton, we take the 
fractional energy density of electron $\epsilon_{e}$ to be $\sim 0.5$ and the 
fractional energy density of the magnetic field $\epsilon_{B}$ to be $\sim 0.01$. 
We integrate equation (5) from 0.5 keV to 8 keV and result in a 
calculated luminosity of $\sim 6\times 10^{32}$ 
ergs s$^{-1}$. With the reasonable choice of parameters stated above, 
the luminosity estimated by this simple model is found to be the same order as 
the observed value.

It is obvious that the radio nebula is significantly longer than its X-ray 
counterpart (cf. Fig. 4). This is not unexpected. Considering a scenario of constant 
injection of particles with a finite synchrotron cooling time, 
the number of particles that can reach at 
a further distance from the pulsar should decrease with increasing frequency. 
This is because the synchrotron cooling timescale decrease with frequency. 
This would result in a fact that the synchrotron nebular size decreases with frequency. 

To further constrain the physical properties of the pulsar wind nebula associated 
with \PSR, multi-wavelength observations are badly needed. Since SUMSS data 
have a rather poor spatial resolution which has a typical beam size of $\sim 45$ arcsec, 
there might be details of the nebular emission remain unresolved. In particular,
it is important to better resolve the nebular emission from the contribution of the SNR 
candidate MSC 319.9-0.7. In view of this, high resolution radio observations 
(e.g. ATCA) are required. In the X-ray regime, although the 
Chandra observation has already provided us with a high resolution image of the nebula, 
the photon statistics is not sufficient to tightly constrain the spectral properties. 
Owing to the superior collecting power, observations with XMM-Newton are expected to put 
a strict constraint on the emission nature of the nebula as well as the pulsar itself. 

Apart from the radio and X-ray observations, a complete study of pulsar wind nebula 
should also include TeV observations (e.g. HESS). It is generally believed that 
the TeV photons are resulted from inverse Compton scattering of soft photon field 
by the relativistic particles in the nebulae. The seed soft photons are possibly 
contributed by the cosmic microwave background (Cui 2006). 
However, there is only a handful of pulsar wind 
nebulae detected in TeV regime so far (see Cui 2006), a larger sample is needed for 
differentiating the aforementioned interpetation from its 
competing scenario (e.g. neutral pion decay).

From the above discussion, one should note that the pulsar's transverse velocity is 
an important parameter in studying the shock physics. And hence measuring the 
proper motion of \PSR\ is badly needed. Moreover, 
although the orientation of the trail suggests \PSR\ is likely moving in the direction 
of northeast, it is not necessary for the trail to be aligned with the pulsar 
velocity. PSRs J2124-3358 and B2224+65 are the examples that the X-ray trails are 
misaligned with the direction of the pulsars' proper motion (Hui \& Becker 2006, 2007).

\end{document}